\documentclass{article}
\usepackage[dvips]{graphicx}

\begin{document}

\title{Analysis of Bifurcations in a Power System Model with Excitation Limits}

\author{ Rajesh G. Kavasseri and K. R. Padiyar \\Deparment of Electrical Engineering \\Indian Institute
of Science, Bangalore, India. \\ Email:
rajesh.kavasseri@ndsu.nodak.edu}
\date{}

\maketitle

\begin{abstract}
\noindent This paper studies bifurcations in a three node power
system when excitation limits are considered. This is done by
approximating the limiter by a smooth function to facilitate
bifurcation analysis. Spectacular qualitative changes in the system
behavior induced by the limiter are illustrated by two case studies.
Period doubling bifurcations and multiple attractors are shown to
result due to the limiter. Detailed numerical simulations are
presented to verify the results and illustrate the nature of the
attractors and solutions involved.
\end{abstract}

\section{Introduction}
Chaos in simple power system models has been studied extensively in
recent papers. In Abed {\em et al.,} [1993], Tan {\em et al.,}
[1993], bifurcations and chaos in a three node power system with a
dynamic load model were studied using a classical model for the
generator. In Rajesh \& Padiyar [1999],  the authors studied dynamic
bifurcations in a similar system and reported the existence of chaos
even with detailed models. However, in Rajesh \& Padiyar [1999],  it
was observed that the field voltage assumed unrealistic values at
the onset of chaos owing to the unmodeled effect of excitation
limits. Though a limiter is fairly easy to model for simulation
purposes, the effect of a limiter on dynamic bifurcations has been
poorly understood because bifurcation analysis demands smoothness of
the functions describing the model. Limit induced chaotic behavior
in a Single Machine Infinite Bus system was studied in Ji and
Venkatasubramanian [1996] by extensive numerical simulations. In
this paper, we approximate the limiter by a smooth function to
facilitate bifurcation analysis and study the changes which arise on
it's consideration. The rest of the paper is organized as follows.
Section 2 deals with the modeling of the system along with the
limiter. Section 3 presents the results of a bifurcation analysis.
Section 4 contains the discussions and Sec. 5, the conclusions.

\section{System Modeling}
\begin{figure}[htbp]
\centerline{\unitlength=0.70mm
\special{em:linewidth 0.4pt}
\linethickness{0.4pt}
\begin{picture}(134.00,65.00)
\put(111.00,54.00){\line(0,-1){17.00}}
\put(111.00,49.00){\line(-1,0){35.00}}
\put(65.00,47.00){\framebox(11.00,4.00)[cc]{}}
\put(65.00,49.00){\line(-1,0){36.00}}
\put(70.00,41.00){\line(0,-1){12.00}}
\put(70.00,38.00){\line(1,0){14.00}}
\put(84.00,36.00){\framebox(11.00,4.00)[cc]{}}
\put(95.00,38.00){\line(1,0){16.00}}
\put(111.00,37.00){\line(0,-1){5.00}}
\put(70.00,38.00){\line(-1,0){14.00}}
\put(45.00,36.00){\framebox(11.00,4.00)[cc]{}}
\put(45.00,38.00){\line(-1,0){16.00}}
\put(29.00,54.00){\line(0,-1){22.00}}
\put(70.00,32.00){\line(1,0){4.00}}
\put(74.00,32.00){\line(0,-1){10.00}}
\put(74.00,12.00){\line(0,-1){9.00}}
\put(72.00,3.00){\line(1,0){4.00}}
\put(67.00,54.00){\makebox(0,0)[cc]{$Y_3\angle{\phi_3}$}}
\put(86.00,42.00){\makebox(0,0)[cc]{$Y_1\angle{\phi_1}$}}
\put(46.00,42.00){\makebox(0,0)[cc]{$Y_2\angle{\phi_2}$}}
\put(84.00,16.00){\makebox(0,0)[cc]{  P,Q}}
\put(111.00,59.00){\makebox(0,0)[cc]{$V_t\angle{\theta}$}}
\put(29.00,58.00){\makebox(0,0)[cc]{$E_b\angle{0}$}}
\put(22.00,65.00){\makebox(0,0)[cc]{Infinite Bus}}
\put(60.00,32.00){\makebox(0,0)[cc]{$V_L\angle{\delta_L}$}}
\put(111.00,43.00){\line(1,0){13.00}}
\put(129.00,43.00){\circle{10.00}}
\bezier{28}(126.00,43.00)(127.00,46.00)(129.00,43.00)
\bezier{28}(129.00,43.00)(130.00,40.00)(132.00,43.00)
\put(70.00,12.00){\rule{8.00\unitlength}{10.00\unitlength}}
\put(106.00,38.00){\vector(-1,0){4.00}}
\put(105.00,49.00){\vector(-1,0){4.00}}
\put(119.00,43.00){\vector(-1,0){4.00}}
\put(115.00,37.00){\makebox(0,0)[cc]{$\hat{i}$}}
\put(100.00,32.00){\makebox(0,0)[cc]{$\hat{i_1}$}}
\put(99.00,54.00){\makebox(0,0)[cc]{$\hat{i_3}$}}
\put(115.00,51.00){\makebox(0,0)[cc]{1}}
\put(65.00,41.00){\makebox(0,0)[cc]{2}}
\put(18.00,45.00){\makebox(0,0)[cb]{3}}
\put(29.00,52.00){\line(-2,-1){4.00}}
\put(25.00,50.00){\line(0,0){0.00}}
\put(29.00,50.00){\line(-2,-1){4.00}}
\put(29.00,48.00){\line(-2,-1){4.00}}
\put(25.00,45.00){\line(0,0){0.00}}
\put(29.00,46.00){\line(-2,-1){4.00}}
\put(29.00,44.00){\line(-2,-1){4.00}}
\put(25.00,42.00){\line(0,0){0.00}}
\put(29.00,42.00){\line(-2,-1){4.00}}
\put(29.00,40.00){\line(-2,-1){4.00}}
\put(25.00,38.00){\line(0,0){0.00}}
\put(29.00,38.00){\line(-2,-1){4.00}}
\put(25.00,36.00){\line(0,0){0.00}}
\put(29.00,36.00){\line(-2,-1){4.00}}
\put(29.00,34.00){\line(-2,-1){4.00}}
\put(25.00,32.00){\line(0,0){0.00}}
\put(29.00,32.00){\line(-2,-1){4.00}}
\put(65.00,41.00){\circle{4.47}}
\put(115.00,51.00){\circle{4.47}}
\put(18.00,46.00){\circle{5.66}}
\put(29.00,54.00){\line(-2,-1){4.00}}
\end{picture}}
\caption{The System}
\label{syst}
\end{figure}
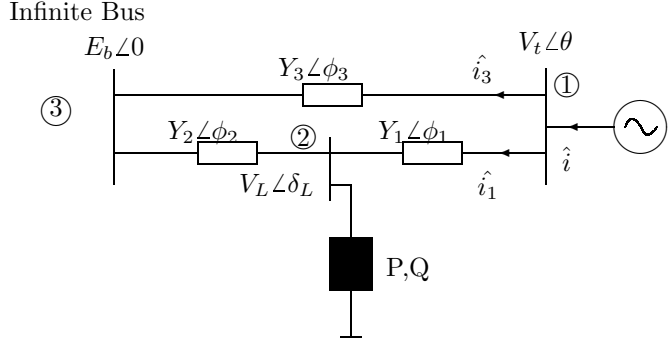

The system as considered in Rajesh \& Padiyar [1999] is shown in Fig.~\ref{syst}.
By a suitable choice of line impedances, we might regard the system as one of a
generator supplying power to a local load which in turn is connected to a remote system
modeled as an infinite bus. For the general reader's convenience, a brief explanation
of the terms d-q and D-Q axis is provided here. The modeling and analysis of three
phase synchronous machines is complicated by the fact that the basic machine equations
are {\bf time varying}. This is circumvented by the use of Park's transformation which
transforms the time varying machine equations in to a time invariant set. The three phase
stator quantities (like voltage, current and flux), when transformed in to Park's frame
yield the corresponding d-q-o variables. When a generator is described in the d-q frame,
then naturally the external network connected to it should also be described in the
same reference frame. However, the non-uniqueness of Park's transformation (each generator
has it's own d-q components) prevents us from doing so. In order to transform the entire
network using a single tranformation with reference to a common reference frame, the
Kron's transformation where the variables are denoted by D-Q-O are used. For a complete
, detailed and clear exposition of these concepts in power system modeling, the reader is
refered to Padiyar [1996].

\subsection{Generator Model}
{\bf Rotor Equations}\\
The rotor mechanical equations for the generator as  given by the swing equations
are,
\begin{eqnarray}
\dot{\delta} = \omega_Bs_m\\
\dot{s_m} = \frac{-ds_m+ P_m - P_g}{2H} \\ \nonumber
\end{eqnarray}
where $d$ is the damping factor in per unit, $\omega_B$ is the system frequency
in rad/s, $P_m$ is the input power to the generator  and  $s_m$, the generator slip defined by   \begin{equation}
s_m = \frac{\omega - \omega_B}{\omega_B} \end{equation}
Two electrical circuits are considered on the rotor, the field winding on
the d-axis and one damper winding on the q-axis. The resulting equations
are,
\begin{eqnarray}
\dot{E^{\prime}_q} = \frac{-E'_q + (x_d-x_d')i_d + E_{fd}}{T^{\prime}_{do}}\\
\dot{E'_d} = \frac{-E'_d - (x_q-x'_q)i_q}{T'_{qo}}\\ \nonumber
\end{eqnarray}
The power delivered by the generator $P_g$ can be expressed as \begin{equation}
P_g = E'_qi_q + E'_di_d + (x'_d - x'_q)i_di_q \end{equation}
{\bf Stator Equations}\\
Neglecting  stator transients and the stator resistance, we have the following
algebraic equations \begin{eqnarray}
E'_q + x'_di_d = v_q\\
E'_d - x'_qi_q = v_d\\ \nonumber
\end{eqnarray}
\subsection {Excitation System }
\begin{figure}[htbp]
\centerline{\unitlength=0.70mm
\special{em:linewidth 0.4pt}
\linethickness{0.4pt}
\begin{picture}(133.00,45.00)(15,15)
\put(47.00,33.00){\circle{8.25}}
\put(26.00,33.00){\vector(1,0){17.00}}
\put(47.00,14.00){\vector(0,1){15.00}}
\put(51.00,33.00){\vector(1,0){15.00}}
\put(66.00,25.00){\framebox(25.00,16.00)[cc]{}}
\put(98.00,37.00){\makebox(0,0)[cc]{$E_{fdx}$}}
\put(47.00,33.00){\makebox(0,0)[cc]{$\Sigma$}}
\put(29.00,37.00){\makebox(0,0)[cc]{$V_{ref}$}}
\put(41.00,36.00){\makebox(0,0)[cc]{+}}
\put(43.00,27.00){\makebox(0,0)[cc]{-}}
\put(50.00,18.00){\makebox(0,0)[cc]{$V_t$}}
\put(78.00,33.00){\makebox(0,0)[cc]{$\displaystyle{\frac {K_A}{1+sT_A}}$}}
\put(123.00,46.00){\makebox(0,0)[cc]{$E_{fd}^{max}$}}
\put(103.00,22.00){\makebox(0,0)[cc]{$E_{fd}^{min}$}}
\put(133.00,29.00){\makebox(0,0)[cc]{$E_{fd}$}}
\put(91.00,33.00){\vector(1,0){39.00}}
\put(115.00,42.00){\line(-1,-3){5.67}}
\put(109.33,25.00){\line(-1,0){7.33}}
\put(122.33,42.00){\line(-1,0){7.33}}
\end{picture}}
\caption{Excitation System}
\label{exci}
\end{figure}
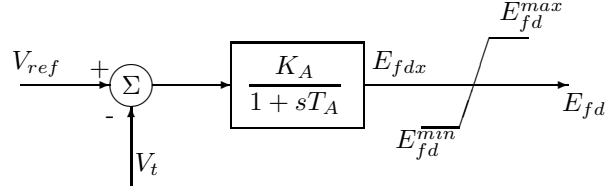

The excitation system for the generator is represented by a single time constant
high gain AVR and the limiter as shown in Fig.~\ref{exci}.

The equation for this excitation system is given by \begin{equation}
\dot{E}_{fdx} = \frac{ -E_{fdx} + K_A(V_{ref} - V_{t})}{T_A}
\end{equation}

\begin{eqnarray}
E_{fd} = E_{fdx} \;if\; E_{fd}^{min} < E_{fdx} < E_{fd}^{max} \label{eq:softi} \\
       = E_{fd}^{min} \;if\; E_{fdx} < E_{fd}^{min} \nonumber \\
       = E_{fd}^{max} \;if\; E_{fdx} > E_{fd}^{max} \nonumber
\end{eqnarray}

\begin{figure}[htbp]
\includegraphics[height=1.5in, width=3in]{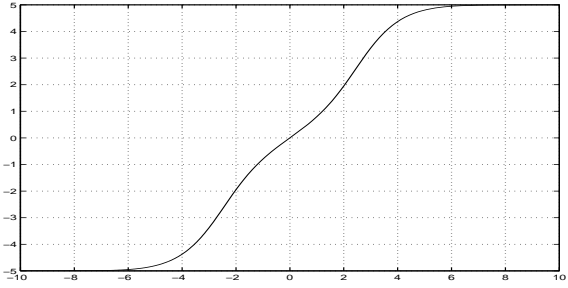}
\caption{The function representing the limiter} \label{flim}
\end{figure}

The limiter shown in Fig.~\ref{exci} and defined by Eq. 10  is a soft
or windup limiter. This limiter model cannot be directly used for bifurcation
studies. An approximate model where the limiter is described by a smooth function
is given below (see Fig.~\ref{flim}). Here, we consider symmetric limits i.e.
$| E_{fd}^{max}|\;=\;| E_{fd}^{min}|\;=\; E_{fdl}$
\begin{equation}
E_{fd} \; = \; f_{lim}(E_{fdx}) \; = \frac{2E_{fdl}}{\pi}tan^{-1}(aE_{fdx}\; exp(bE^2_{fdx}))
\end{equation}

{\bf Remarks}\\
Such an approximation amounts to perturbing the vector field slightly and
hence the equilibrium structure of the system will also be slightly perturbed.
So in our studies, the focus will be on how the limiter influences
non-stationary solutions and their bifurcations.

\subsection{Load Model}
A dynamic load model as in Abed [1993] is used along with a constant power load
$ (P_{ld},Q_{ld}) $ in parallel with it. Thus, the real and reactive load powers
are specified by the following equations.
\begin{eqnarray}
P &=& P_{ld} + P_o + p_1\dot{\delta_L} + p_2\dot{V_L} + p_3V_L \label{eq:pld}\\
Q &=& Q_{ld} + Q_o + q_1\dot{\delta_L} + q_2V_L + q_3V_L^2 \label{eq:qld}
\end{eqnarray}

\subsection{Network Model}
With the notation defined in Fig. 1, we can write the network equation in the D-Q reference
frame as,
\begin{eqnarray}
\hat{E_b} + \frac{\hat{i_3}}{\hat{Y_3}} = \hat{V_t} \label{eq:fir}\\
\hat{V_L} + \frac{\hat{i_1}}{\hat{Y_1}}=\hat{V_t}
\end{eqnarray}
\newline
Further,
\begin{eqnarray}
\hat{V_t} \;\;\; = (v_q +jv_d)e^{j\delta} \label{eq:vt} \\
\hat{i}\;\;\; =\;\;\; (i_q+ji_d )e^{j\delta}\;\;\;=\;\;\;\hat{i_1}\;\;\; +\;\;\; \hat{i_3}\\
\hat{Y} =  Y\angle{\phi}  = \hat{Y_1} + \hat{Y_3} \label{eq:las}
\end{eqnarray}
From Eqs., (\ref{eq:fir}) to (\ref{eq:las}) we can write,
\begin{equation}
(v_q+jv_d) = \frac{A_1 + B_1}
{Y}
\end{equation}
where $A_1=E_bY_3e^{-j(\delta+\phi - \phi_3)}+
Y_1V_Le^{j(\delta_L-\delta-\phi +\phi_1)}$ \\and $B_1=(i_q+ji_d)e^{-j\phi}$\\
Defining,
\begin{eqnarray}
a = E_bY_3cos(\delta+\phi-\phi_3) + Y_1V_Lcos(\delta_L-\delta-\phi+\phi_1)\\
b = -E_bY_3sin(\delta+\phi-\phi_3) + Y_1V_Lsin(\delta_L-\delta-\phi+\phi_1)
\end{eqnarray}
\newline
permits us to write,
\begin{eqnarray}
i_qcos(\phi) + i_dsin(\phi) = Yv_q - a\\
i_dcos(\phi) - i_qsin(\phi) = Yv_d -b
\end{eqnarray}

\subsection {Derivation of the System Model}

Substituting for $v_d$ and $v_q$ from the stator algebraic equations (7) and (8), we have,\\
\begin{eqnarray}
\left[\begin{array}{cc}
cos(\phi) & (sin(\phi)-Yx^{\prime}_d)\\
-(sin(\phi)-Yx^{\prime}_q) & cos(\phi)
\end{array} \right]\left[\begin{array}{c} i_q \\ i_d \end{array}\right]
=\left[\begin{array}{c} Y_a\\Y_b\end{array}\right]\\  \nonumber
\label{eq:sys}
\end{eqnarray}
where \hspace*{0.1cm} $Y_a=(YE^{\prime}_q -a)$\\
and \hspace*{0.3cm} $Y_b=(YE'_d -b)$

From Eq. (24), we can solve for the currents $i_d,\;i_q$
and subsequently solve for $ v_d\;$and$\;v_q$ from the stator
algebraic equations. Further, from Eq. (\ref{eq:vt}) we get,\begin{eqnarray}
|{\hat{V_t}}| =\sqrt{(v_q^2 + v_d^2)}\\
\theta = \delta + tan^{-1}({\frac{v_d}{v_q}}) \label{eq:thet}
\end{eqnarray}
Defining,\begin{eqnarray}
r_1\;\;\;=\;\;\; \delta_L - \theta - \phi_1\\
r_2 \;\;\;=\;\;\;\delta_L-\phi_2\\  \nonumber
\end{eqnarray}
the power balance equation at bus 2 can be written as,
\begin{eqnarray}
P =  V_tV_LY_1cos(r_1) - V_L^2Y_1cos(\phi_1) +  E_bV_LY_2cos(r_2) \nonumber \\-  V_L^2Y_2cos(\phi_2)\label{eq:p}\\
Q =   V_tV_LY_1sin(r_1) + V_L^2Y_1sin(\phi_1) +  E_bV_LY_2sin(r_2) \nonumber \\+  V_L^2Y_2sin(\phi_2)\label{eq:q}
\end{eqnarray}
Substituting from Eqs., (\ref{eq:sys}-\ref{eq:thet}), (\ref{eq:p}-\ref{eq:q}) in Eqs., (1-5) and (\ref{eq:pld}-\ref{eq:qld}), we get
\begin{eqnarray}
\bf \dot{x} = \bf f(\bf x,\lambda)
\end{eqnarray} where
\begin{math}
\bf x =
\left[\begin{array}{ccccccc}
\delta & s_m &  E'_q & E'_d & E_{fdx}& \delta_L & V_L\end{array}\right]^T\end{math}
and $\lambda$ is a  bifurcation parameter. As a simplification, we shall also
consider the system described the One Axis Model for the generator as the
effect of the limiter on this case is interesting in itself. For this, we neglect
the damper winding on the q-axis and in terms of modeling, this is done by
omitting  $E'_d$ as a state variable and substituting
\begin{equation}
E'_d = -(x_q - x'_q)i_q \end{equation}
in Eqs., (6) and (8). The
state space structure remains the same, with the dimension being one less
that the previous system. In this case, we have
\begin{math}
\bf x =
\left[\begin{array}{cccccc}
\delta & s_m &  E'_q  & E_{fdx}& \delta_L & V_L\end{array}\right]^T\end{math}.

\section{Bifurcations}
In this section, we illustrate the qualitative differences which arise on consideration
of the limiter by studying bifurcations in the associated systems with {\bf AUTO97} (Doedel [1997])
a continuation and bifurcation software for ordinary differential equations. The generator
input power ($P_m$) is a very important parameter in practical power systems operation. This is
the parameter which is adjusted or varied by the power system operators (utility) to track the
changes and variations in the system load (power demand) so as to maintain a stable operating condition.
We hence, consider $P_m$ i.e.
the input power to the generator as the bifurcation parameter. To describe the types of bifurcations, we shall use the following notations.\newline
SNB:  Saddle Node Bifurcation\\
HB:   Hopf Bifurcation\\
CFB: Cyclic Fold Bifurcation\\
TR : Torus Bifurcation\\
PDB : Period Doubling Bifurcation\\
In all the bifurcation diagrams
the state variable $E_{fdx}$ is plotted against the bifurcation parameter. In the case
of periodic solutions, we use the maximum value of the variable which is indicated by the
circles. Filled circles refer to stable solutions and the unfilled ones, to unstable solutions.

\subsection{One Axis Model}
\begin{figure}
\includegraphics[height=2in, width=3in]{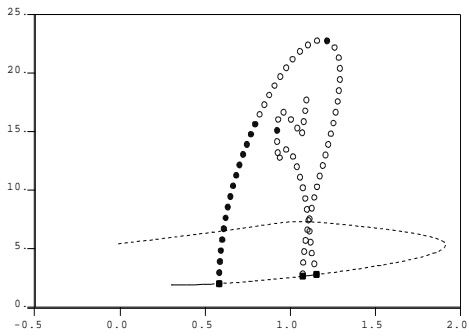}
\caption{\scriptsize{$\lambda = P_m$, One Axis Model}} \label{sax1}
\end{figure}

\begin{figure}
\includegraphics[height=2in, width=3in]{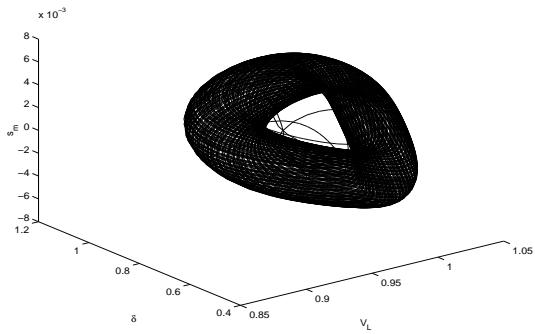}
\caption{\scriptsize{The Quasi-periodic trajectory when $P_m =
0.83$}} \label{mas1}
\end{figure}

{\bf Without limiter}\\
From Fig.~\ref{sax1}, we note that the stationary solutions
undergo four bifurcations labeled as HB$^1$, HB$^2$, HB$^3$ and SNB$^4$.
For $\lambda < \lambda^1$, the equilibrium point is stable, but as $\lambda$ is increased,
the stationary point loses its stability at $\lambda$= $\lambda^1$ through HB$^1$. With a
further increase in  $\lambda$, the stationary point gains stability through HB$^2$, i.e.
$\lambda$= $\lambda^2$. It remains stable until $\lambda$= $\lambda^3$, where stability
is lost through HB$^3$. Further, SNB$^4$ does not influence the stability of the stationary
point. Next, we focus on the family of periodic solutions emerging from HB$^1$.
Since HB$^1$ is
supercritical, it gives birth to a family of stable periodic solutions indicated by the
filled circles. This periodic solution loses its stability at TR$^5$ and with a further increase
in  $\lambda$, gains it back through TR$^6$ and remains stable until TR$^7$. Further on, there is
no qualitative change in its behavior with TR$^7$, CFB$^8$ and TR$^9$. Next, we find
that the branch
emerging on continuation of HB$^2$ is the same as that from HB$^1$.
On continuation of HB$^3$, we find a family of unstable periodic solutions which gain stability
through CFB$^{10}$. This stable periodic solution encounters a PDB$^{11}$
on continuation
of which, we find a period doubling cascade accumulating at a critical value of  $\lambda^c$ =0.931.
(which is not shown here)
which definitely suggests the onset of chaos.
However, what is of interest here, is the behavior of the system after TR$^5$. It is clear that a
torus bifurcation results in the emergence of quasi-periodic solutions. This is verified by simulation
as shown in Fig.~\ref{mas1} which shows the quasi-periodic attractor.
The bifurcation points are summarized in Table ~\ref{tsax1}
\begin{small}
\begin{table}[htbp]
\begin{center}
\caption{Bifurcation Points (see Fig.~\ref{sax1}) \label{tsax1} }
\begin{tabular}{|l|l|l|l|l|l|l|l|l|l|l|l|} \hline
Point&HB$^1$&HB$^2$ &HB$^3$&SNB$^4$&TR$^5$&TR$^6$&TR$^7$&
CFB$^8$&TR$^9$&CFB$^{10}$&PDB$^{11}$ \\ \hline
$\lambda$ &0.583&1.0746&1.155&1.914&0.812&1.181&1.26&1.293&1.26&0.922&0.9278\\ \hline
\end{tabular}
\end{center}
\end{table}
\end{small}

\begin{figure}[htbp]
\includegraphics[height=2in, width=3in]{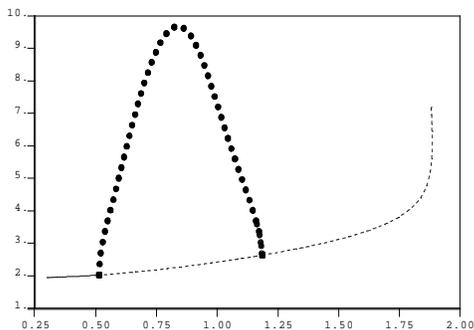}
\caption{$\lambda = P_m$ , One  Axis Model with limiter,continuation
of HB$^1$ and HB$^2$} \label{lima9}
\end{figure}

\begin{figure}[htbp]
\includegraphics[height=2in, width=3in]{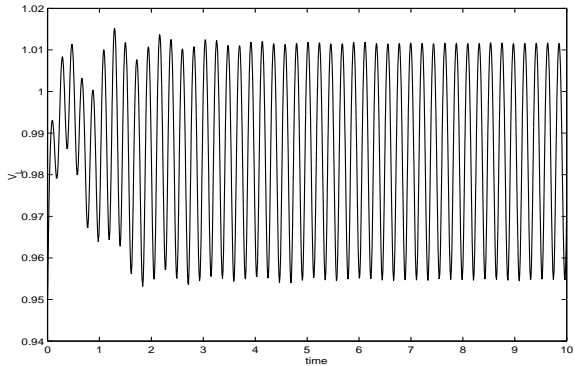}
\caption{Sustained  oscillations of load bus voltage with time when
$\lambda = 0.86$ with the approximate limiter} \label{mal13}
\end{figure}

{\bf With Limiter}\\
From Fig.~\ref{lima9}, we observe that the stable operating point loses its
stability with HB$^1$, regains it at HB$^2$ and loses it back at HB$^3$ before
encountering SNB$^4$ which is similar to the case without limiter (see Fig.~\ref{sax1}).
Note that in Fig.~\ref{sax1}, for the static bifurcations HB$^1$ - SNB$^4$,
$E_{fdx} < E_{fd}^{max}$ and hence we expect that these bifurcations should
occur at the same values even with the limiter.
However, this is not the case as seen from Table ~\ref{t42} because of the
approximation which shifts the equilibrium structure as mentioned before.
HB$^2$ and HB$^3$  occur very  closely and hence cannot be distinguished in Fig.~\ref{lima9}.
On continuation of HB$^1$ which is supercritical, we find that the stable periodic
solutions do not undergo any bifurcation. HB$^2$ is also supercritical and its
continuation yields the same stable periodic set obtained on continuation of
HB$^1$. HB$^3$ is sub-critical and its continuation which yields CFB$^5$ where
stability is gained for a while before CFB$^6$ is however, not shown here.
Fig.~\ref{mal13} shows the time domain plot of the
load bus voltage for $\lambda = 0.86$.
The bifurcations are
summarized in Table ~\ref{t42}.
\begin{small}
\begin{table}[htbp]
\begin{center}
\caption{Bifurcation Points (see Fig.~\ref{lima9})
\label{t42}}
\begin{tabular}{|l|l|l|l|l|l|l|} \hline
Point&HB$^1$ & HB$^2$ & HB$^3$ & SNB$^4$&CFB$^5$&CFB$^6$\\ \hline
$\lambda$ & 0.5145 &1.1827 &1.1857&1.884&1.1284&1.1311\\ \hline
\end{tabular}
\end{center}
\end{table}
\end{small}

\subsection{Two Axis model}
\begin{figure}[htbp]
\includegraphics[height=2in, width=3in]{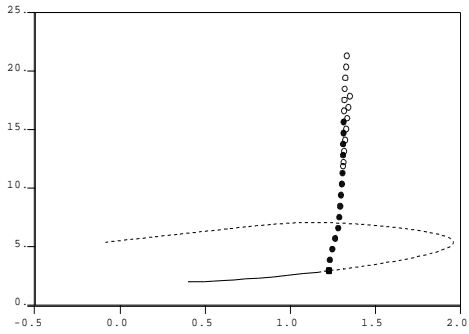}
\caption{\scriptsize{$\lambda = P_m$, Two Axis Model}} \label{sax2}
\end{figure}

\begin{figure}[htbp]
\includegraphics[height=2in, width=3in]{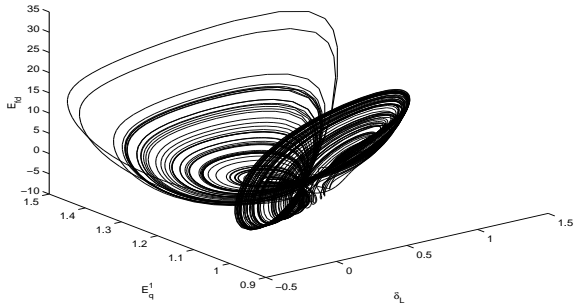}
\caption{\scriptsize{The Chaotic trajectory when $P_m$ = 1.315}}
\label{mas2}
\end{figure}

{\bf Without limiter}\\
We let  $\lambda= P_m$ with reference to Fig.~\ref{sax2}. The stationary point undergoes
two bifurcations, HB$^1$  where it loses its stability and SNB$^2$
which does not influence the stability further. HB$^1$ is a supercritical
bifurcation and the family of stable periodic solutions from it undergo a period doubling
cascade starting with the PDB$^1$, accumulating at a critical value
of  $\lambda^c= 1.315$. The chaotic attractor at  $\lambda^c$
is shown in Fig.~\ref{mas2}
which confirms the chaotic nature. The bifurcation points are summarized in Table~\ref{tsax2}
\begin{small}
\begin{table}[htbp]
\begin{center}
\caption{Bifurcation Points (see Fig.~\ref{sax2}) \label{tsax2} }
\begin{tabular}{|l|l|l|l|l|} \hline
Point&HB$^1$& SNB$^2$&PDB$^3$&PDB$^4$  \\ \hline
$\lambda$ & 1.2281&1.9607&1.311&1.314 \\ \hline
\end{tabular}
\end{center}
\end{table}
\end{small}

\begin{figure}[htbp]
\includegraphics[height=2in, width=3in]{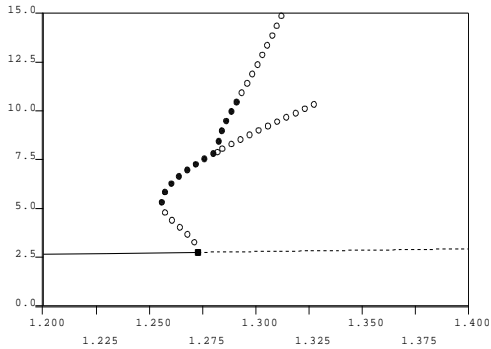}
\caption{$\lambda = P_m$, Two Axis Model with limiter} \label{lima1}
\end{figure}

\begin{figure}[htbp]
\includegraphics[height=2in, width=3in]{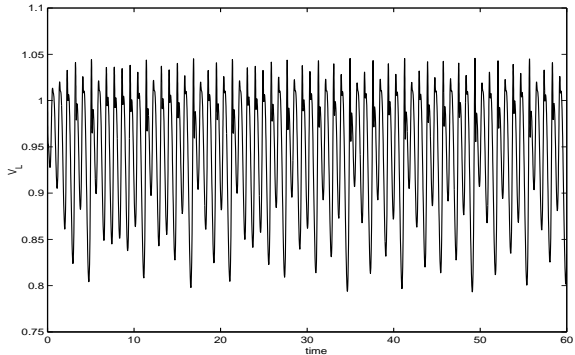}
\caption{Chaotic oscillations of load bus voltage with time when
$\lambda = 1.3$ with the approximate limiter} \label{ex5}
\end{figure}

\begin{figure}[htbp]
\includegraphics[height=2in, width=3in]{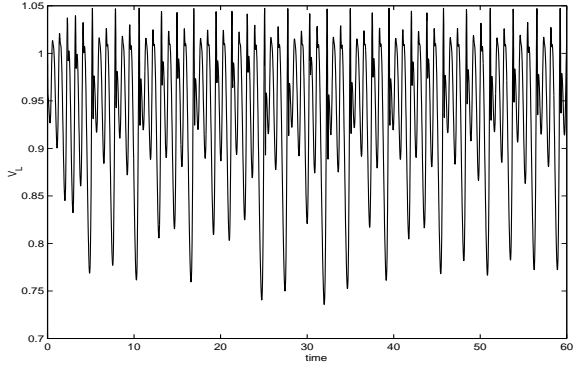}
\caption{Chaotic oscillations of load bus voltage with time when
$\lambda = 1.3$ with the exact limiter} \label{ex6}
\end{figure}

\begin{figure}[htbp]
\includegraphics[height=2in, width=3in]{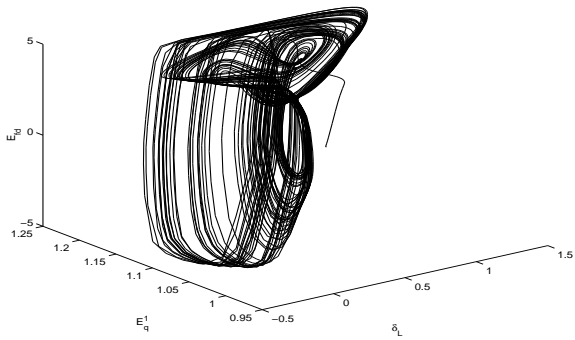}
\caption{The Chaotic attractor subject to limits when $\lambda =
1.3$ } \label{ex12}
\end{figure}

{\bf With Limiter}\\
From Fig.~\ref{lima1}, we observe that the stable operating point loses
stability at HB$^1$ and then encounters SNB$^2$ (which is not shown here).
On continuation of HB$^1$, which
is sub-critical, we find that the unstable periodic solution stabilizes with
CFB$^3$. This stable periodic solution undergoes a period doubling cascade
initiated at PDB$^4$. In Fig.~\ref{lima1},we also show the period doubled
solution and its subsequent bifurcation PDB$^5$. By numerical simulations,
considering both the exact and the function approximation of the limiter,
we verify that at  $\lambda =1.3$, the system
behavior is chaotic. The time domain plots are shown in Figs ~\ref{ex5}
and ~\ref{ex6}. The chaotic attractor subject to limits is shown in Fig ~\ref{ex12}.
The bifurcations are summarized in Table ~\ref{t41}.
\begin{small}
\begin{table}[htbp]
\begin{center}
\caption{Bifurcation Points (see Fig.~\ref{lima1})
\label{t41}}
\begin{tabular}{|l|l|l|l|l|l|} \hline
Point&HB$^1$ & SNB$^2$ & CFB$^3$ & PDB$^4$&PDB$^5$\\ \hline
$\lambda$ & 1.2729 &1.923 &1.2557&1.282&1.2912\\ \hline
\end{tabular}
\end{center}
\end{table}
\end{small}

\section{Discussions}
The case studies with two different models were considered solely for
illustrating the effect of the limiter on bifurcations in the system which is interesting
in it's own right. The qualitative difference in system dynamics owing to modeling is however
not discussed here (see Rajesh and Padiyar [1999] for a discussion). Another aspect
worth mentioning is the differences in the bifurcation diagrams in this paper from those
in the references. Abed {\em et al} consider a simplified generator model (classical model)
in which the excitation system is entirely absent and use a slightly different system for the
bifurcation studies. In Ji and Venkatasubramanian [1996], a Single Machine Infinite Bus
(SMIB) system (which is different from that considered in this paper) wherein the
load model is absent, is studied. This paper however, focusses mainly on studying
bifurcations and changes which arise on the consideration of {\bf excitation limits}.
When the One axis model is considered
without the limiter, we observe stable quasi-periodic trajectories resulting from a TR
bifurcation. However, with the limiter, we do not observe any bifurcations of periodic
solutions with the result that the entire branch from HB$^1$ to  HB$^2$ in Fig. is stable.
When the Two axis model is considered without the limiter, we observe chaotic trajectories
due to PDBs, which, with the limiter still occur. However, we observe in this case that
the system has multiple attractors (see Fig.~\ref{lima1}) namely, a stable equilibrium point and
a stable periodic solution. Further, we observe that the PDBs in this case occur very
close to the boundary of stable fixed point operation.
This means that if the system operates close to boundary of stable fixed point operation, and
suffers a disturbance with the post-disturbance initial condition belonging to the chaotic
region, the system can be easily pushed to the chaotic region. Another interesting aspect
seen by comparing Fig.~\ref{sax2} and Fig.~\ref{lima1} is that stable equilibrium points close
to the boundary of stable fixed point operation are surrounded by unstable limit cycles
which suggests that the region of attraction for the equilibrium points {\em shrinks} in
the presence of limits.

\section{Conclusions}
An attempt has been made to analyze bifurcations in the presence of a limiter by approximating
the limiter by a smooth function. It is seen that this methodology provides good insight
in to studying bifurcations in a system with a soft limiter. The observations in the case studies
illustrate in general that, the limiter is capable of inducing spectacular qualitative changes
in the system. Developing a formal theory for bifurcations and analyzing the global system dynamics
in the presence of limits in the system would be a challenging task for further research in
this area.

\newpage

{\bf References} \\

\noindent  Abed E. H., Wang H. O., Alexander J. C., Hamdan A. M. A. and Lee H. C. [1993] ``Dynamic bifurcations
in a power system model exhibiting voltage collapse'', {\em Int. J. Bifurcations and Chaos} {\bf 3} (5),
1169-1176. \\

\noindent Doedel E. J., Fairgrieve T. F., Sanstede B., Champneys A. R., Kuznetsov and Wang X [1997] ``AUTO97 :
(User's Manual) continuation and bifurcation software for ordinary differential equations.'' \\

\noindent Ji W. and Venkatasubramaian V [1996] ``Hard limit induced chaos in a fundamental power system model'',
{\em International Journal of Electrical Power and Energy Systems} {\bf 18} (5), 279-295.\\

\noindent Padiyar K. R. [1996] ``Power System Dynamics and Control'' John Wiley, Singapore. \\

\noindent Rajesh K. G. and Padiyar K. R. [1999] ``Bifurcation analysis of a three node power system with detailed
models'' {\em International Journal of Electrical Power and Energy Systems} {\bf 21}  375-393.\\

\noindent Tan C. W., Varghese M., Varaiya P. P. and Wu F. W.[1995]  ``Bifurcation,chaos and voltage collapse in power systems''
{\em Proceedings of the IEEE} {\bf 83} (11) 1484-1496.

\

\section*{Appendix A}

\begin{itemize}
\item Network parameters.
\newline
\begin{math}
Y_1 = 4.9752,  Y_2 = 1.6584,  Y_3 = 0,\phi_1 = \phi_2=\phi_3 = -1.4711
,E_b = 1.0\end{math}
\item Generator parameters.
\newline
\begin{math}
x_d = 1.79, \;\;\;x_q = 1.71, \;\;\;T'_{do}=4.3, T^{\prime}_{qo}=0.85, x_d'=0.169,
\newline
 x_q'=0.23, H = 2.894,w_b = 377,d = 0.05 ,E_m = 1.0
\end{math}
\item Load parameters.
\newline
\begin{math}
P_o = 0.4,\;\;,Q_o = 0.8\;\;,p_1 = 0.24\;\;,q_1 = -0.02\;\;,p_2 = 1.7\;\;,q_2 = -1.866
\newline
p_3 = 0.2\;\;,q_3 = 1.4\;\;
\end{math}
\item AVR constants
\newline
\begin{math}
K_A = 200\;\;,T_A = 0.05
\end{math}
\item Limiter constants
$a = 0.23\;,\;b = 0.1058\;,\;E_{fd}^{max} = 5\;,\;E_{fd}^{min} = -5$
\end{itemize}
\end{document}